\documentclass[twocolumn,showpacs,pre,aps]{revtex4}

\usepackage{dcolumn}
\usepackage{bm}

\begin{document}

\title{On Statistics and 1/f Noise of Brownian Motion\\
in Boltzmann-Grad Gas and Finite Gas on Torus. I. Infinite Gas}

\author{Yuriy E. Kuzovlev}
\email{kuzovlev@kinetic.ac.donetsk.ua} \affiliation{Donetsk Institute
for Physics and Technology, 83114 Donetsk, Ukraine}


\begin{abstract}
An attempt is made to compare statistical properties of
self-diffusion of particles constituting gases in infinite volume and
on torus. In this first part, equations are derived which represent
roughened but solvable variant of the collisional approximation to
exact BBGKY equations. With their help, statistics of Brownian motion
in infinite gas is considered, under the Boltzmann-Grad limit, and
shown to be essentially non-Gaussian, involving 1/f fluctuations in
diffusivity.
\end{abstract}


\maketitle

\section{Introduction}

In \cite{i1} (see also Part VII and then in \cite{i2}) it was argued
that kinetics of statistically nonuniform (though may be
thermodynamically equilibrium) gas does not reduces to the Boltzmann
equation, even under the Boltzmann-Grad limit
($\,\nu\rightarrow\infty\,$, $\,\delta\rightarrow 0\,$,
$\,\nu\delta^2=\,$const\,, with $\,\nu\,$ being concentration of gas
particles and $\,\delta\,$ radius of their repulsive interaction).
Moreover, just this limit makes clear why the Boltzmann's
``Stosshalansatz'' (``molecular chaos'' hypothesis) fails and the
Boltzmann equation becomes inadequate when the one-particle
distribution function $\,F_1(t,\bm{r},\bm{v})\,$ really depends on
spatial coordinate $\,\bm{r}\,$. Instead, an infinite chain of
kinetic equations (reproduced in Appendix B) arises, which involves,
in addition to $\,F_1\,$, specific $\,n$-particle distribution
functions $\,F_n(t,\bm{r},\bm{v}_1,...,\bm{v}_n)\,$ representing
local (probability) densities of $\,n$-particle chains of collisions
(strictly speaking, impact parameters and/or other internal
parameters of the collisions also are arguments of the $F_n$). The
resulting long-scale statistics of self-diffusion (Brownian motion)
of gas particles displays low-frequency fluctuations (1/f\,-noise) of
diffusivity, which correspond to not the Wiener process but
essentially different, non-Gaussian and non-Marcovian, random walk
(thus confirming earlier phenomenological considerations
\cite{bk1,bk2,bk3}).

In contrary to that conjectures, the mathematical theory of dynamical
systems with ``hyperbolic behavior'' (exponential instability and
mixing of phase trajectories), in particular, such as $\,N\,$ hard
balls in a container or on torus (most interesting for me), succeeds
in proof of various ergodic, Bernoullian and Marcovian properties of
these systems (see e.g. \cite{an,cy,ns} and references therein). In
respect to hard balls on torus, this implies, undoubtedly, that their
Brownian motion along it reduces (at sufficiently rough time scales)
to the usual Wiener process (one can easy make sure of this by
numeric simulation of 3 hard balls on 2-D torus).

It would be interesting to reconcile these rigorous results with
above mentioned non-rigorous results on the Boltzmann-Grad gas (which
can be placed on torus as well). There is no contradiction, in
principle, if maximal time scale, $\,\tau_{ng}\,$, which still
perceives ``anomalous'' non-Gaussian and non-Marcovian features of
Brownian paths in $\,N$-particle system (or minimum time scale which
still suppresses and loses these features), is a growing function of
$\,N\,$, for instance,
\begin{equation}
\frac {\tau_{ng}(N)}{\tau_f(N)}\,\sim\,N\,\,,\label{tng}
\end{equation}
where $\,\tau_f(N)\,$ is mean free flight time (we take in mind that
$\,1\ll N<\infty\,$, of course).

I can suggest heuristic arguments in favour of this hypothesis.
Indeed, consider Brownian path of some particle during some time
interval $\,\Delta t\,$. The corresponding fragment of the path
consists of $\,\sim n=\Delta t/\tau_f\,$ collisions and free flights
and is fully characterized by $\,N_d\sim 2\mathcal{D}n\,$ details
($\,\mathcal{D}\,$ denotes torus dimension), namely, velocity vectors
$\,\bm{v}_j\,$ and flight times $\,\tau_j\,$ ($\,j=1\div n\,$) and
besides $\,(\mathcal{D}-1)$-dimensional impact parameters of
collisions. Compare this quantity with a number $\,N_c\,$ of initial
conditions (parameters of the gas state at the beginning of the
interval) which might have an influence on this fragment. It is
natural to expect that statistics of the Brownian displacement
$\,\Delta R_s=\bm{v}_1\tau_1+...+\bm{v}_s\tau_s\,$ ($\,s=1\div n\,$)
is sensitive to the ratio $\,N_d/N_c\,$. If $\,N_d/N_c\ll 1\,$ then a
particular realization of $\,\Delta R_s\,$ is hardly able to try all
principal possibilities allowed by initial conditions, therefore, it
can be essentially non-ergodic, in one or another sense. In opposite,
if $\,N_d/N_c\gg 1\,$ then any of qualitatively different
possibilities will be sampled many times, which gives grounds for
good ergodicity.

To estimate $\,N_c\,$, notice that outside particle coming to
$\,j$-th collision ($\,j=1\div n\,$) previously had $\,\sim j\,$
collisions with other particles, which in their turn also had
collisions before, and so on. Therefore total number, $\,N_p\,$, of
particles whose initial states influence our fragment can be
estimated as $\,N_p\sim (n+1)^{n+1}/(n+1)!\sim e^n/\sqrt{n}\,$. At
$\,\Delta t/\tau_f =n >\,\ln\,N\,$, in fact, initial state of the
whole gas is important, $\,N_p=N\,$. Correspondingly,
$\,N_c=2\mathcal{D}N_p\sim 2\mathcal{D}\min(N,e^n/\sqrt{n})\,$.

In case of the Boltzmann-Grad gas $\,N=\infty\,$, and $\,N_d/N_c\sim
n^{3/2}e^{-n}\,$ is certainly very small. Hence, non-ergodicity of
$\,\Delta R_n\,$ (more precisely, ``anomalous'' behavior of fourth
and higher-order cumulants of $\,\Delta R_n\,$ \cite{i1,i2}) is not
surprising. At finite $\,N\,$ (and not too short time intervals,
$\,\Delta t >\tau_f\,\ln\,N\,$), $\,N_d/N_c=\Delta t/N\tau_f\,$.
Hence, the border between ergodic and non-ergodic behavior lies at
$\,\Delta t/\tau_f \sim N\,$. That is nothing but our hypothesis
(\ref{tng}).

In view of above reasonings, a system of large finite number of
particles on torus is of special interest. Unfortunately, even in the
much simpler case of infinite Boltzmann-Grad gas the chain of kinetic
equations discussed in \cite{i1,i2} seems not solvable somehow except
their more or less caricature roughening. Therefore, in the present
paper I confine myself by two unpretentious tasks. In this first part
of the paper, the mentioned equations are subjected to even more
grotesque simplification but allowing for their formally exact
solution. At that, their principal structure is preserved, therefore
the solution is must be quite meaningful. In the next second part,
modification of the roughened equations for finite $\,N\,$ will be
derived and adapted to torus. Their analysis indeed demonstrates
existence of characteristic time $\,\tau_{ng}\,$ which both formally
and essentially corresponds to (\ref{tng}). Other results will be
commented in final Conclusion.

\section{Equations of self-diffusion in infinite gas}
Our consideration will be based on equations of the collisional
approximation (see Appendix B and \cite{i1,i2}), first of all, on
equations (\ref{lo}) adapted for thermodynamically equilibrium but
statistically non-uniform gas when Boltzmannian collision operator is
replaced by the Boltzmann-Lorentz collision operator \cite{rdl}.

Recall that equations (\ref{lo}) deal with $\,n$-particle
distribution functions $\,F_n(t,\bm{R},\bm{v}_1...\bm{v}_n)\,$ which
describe close encounters of $\,n\,$ particles (with pair collisions
between some of them) clustered in vicinity of given point
$\,\bm{R}\,$. The closeness means that the particles belong to a
``collision box'' defined by $\,|\bm{r}_j-\bm{R}|<d\,$, where $\,d\,$
is suitable scale much greater than interaction radius $\,\delta\,$
but mush smaller than dimension $\,\sim (n/\nu)^{1/\mathcal{D}}\,$ of
volume typically needed to contain $\,n\,$ particles ($\,\nu\,$ and
$\,\mathcal{D}\,$ are mean number of particles per unit volume and
space dimension of gas, respectively). At that, the general position
$\,n$-particle distribution functions (DF)
$\,F_n(t,\bm{r}_1...\bm{r}_n,\bm{v}_1...\bm{v}_n)\,$, for particles
out of any encounter, are factored into product $\,\prod_{j=1}^n
F_n(t,\bm{r}_j,\bm{v}_j)\,$, at least, under the Boltzmann-Grad limit
$\,\nu\rightarrow\infty\,$, $\,\delta\rightarrow 0\,$, $\,\lambda\sim
1/\nu\delta^{\mathcal{D}-1} =\,$const\,, where $\,\lambda\,$ is mean
free path. These DF are normalized to gas volume $\,\Omega\,$:
$\,\int F_n(t,\bm{r}_1...\bm{r}_n,\bm{v}_1...\bm{v}_n)\,
d\bm{v}\,d\bm{r}=\Omega^n\,$. Below it will be convenient to name
$\,F_n(t,\bm{R},\bm{v}_1...\bm{v}_n)\,$ cluster distribution
functions (CDF) while
$\,F_n(t,\bm{r}_1...\bm{r}_n,\bm{v}_1...\bm{v}_n)\,$ volume
distribution functions (VDF).

The stationary solution to the equations (\ref{lo}) is
\[
F_n(\bm{R},\bm{v}_1...\bm{v}_n)\,=\,\prod_{j=1}^n
F_0(\bm{v}_j)\,\equiv \,F_n^{\,eq}(\bm{v})\,\,\,,
\]
where $\,F_0(\bm{v})=(2\pi
v_0^2)^{-\mathcal{D}/2}$$\exp{(-\bm{v}^2/2v_0^2)}\,$ is equilibrium
velocity distribution, $\,v_0^2=T/M\,$, and $\,T\,$ is gas
temperature. Such DF represent statistical ensemble which tells
nothing about positions of particles.

Now, consider more interesting ensemble which tells us that one of
particles is localized, say, nearby coordinate origin, with
coordinate probability distribution $\,W(\bm{r})\,$ ($\,\int
W(\bm{r})\,d\bm{r}\,=1\,$, $\,\int \bm{r}W(\bm{r})\,d\bm{r}\,=0\,$).
To write this information on VDF, at first suppose that gas volume
and total number of particles $\,N=\nu\Omega\,$ are finite. Then the
highest VDF can be written as
\[
F_N\,=\,F_N^{\,eq}(\bm{v})\,\frac {\Omega}{N}\,\sum_{j=1}^N
\,W(\bm{r}_j)
\]
According to relation $F_n=\int
F_{n+1}\,d\bm{v}_{n+1}d\bm{r}_{n+1}/\Omega\, $, the corresponding
lower VDF are
\begin{equation}
F_n\,=\,F_n^{\,eq}(\bm{v})\left(1+\frac 1\nu\,\sum_{j=1}^n
\left[W(\bm{r}_j)-\frac 1 {\Omega}\right ]\right )\,\label{op}
\end{equation}
Tending $\,\Omega\,$ to infinity and choosing the distribution
$\,W(\bm{r})\,$ to have a suitable width (e.g. $\,\sim \lambda\,$),
it is easy to see that corresponding CDF can be approximately
represented by
\begin{equation}
F_n(\bm{R},\bm{v}_1...\bm{v}_n)\,=\,F_n^{\,eq}(\bm{v})\left[\,1+\frac
n\nu\,\,W(\bm{R})\right ] \,\label{cop}
\end{equation}
In the Boltzmann-Grad limit (when size of any ``collision box'' turns
into zero in units of $\,\lambda\,$) this is formally exact
expression.

Hence, in order to investigate self-diffusion of initially localized
particle, we should solve equations (\ref{lo}) starting from
statistically (but not thermodynamically) non-equilibrium initial
conditions presented by (\ref{cop}). Obviously, contribution of the
first term on right-hand side of (\ref{cop}), that is uniform
background of densities of particles and clusters, will be
time-independent. Therefore we will omit it from solution, together
with constant multiplier $\,1/\nu\,$ from its time-dependent part,
but keep designation $\,F_n\,$ for the rest of it. Thus we come to
the problem
\begin{equation}
\frac {\partial F_n}{\partial t}\,=\,-\bm{V}^{(n)}\cdot\frac
{\partial F_n}{\partial \bm{R}}+\nu\sum_{j=1}^n \Lambda_j\int
F_{n+1}\,d\bm{v}_{n+1}\, \label{lod}
\end{equation}
for infinite set of functions $\,F_n(t,\bm{R},\bm{v}_1...\bm{v}_n)\,$
under initial conditions
\begin{equation}
F_n(t=0,\bm{R},\bm{v}_1...\bm{v}_n)\,=\,n\,W(\bm{R})\prod_{j=1}^n
F_0(\bm{v}_j) \,\label{is}
\end{equation}
with $\,\bm{V}^{(n)}=(\bm{v}_1+...+\bm{v}_n)/n\,$ and $\,\Lambda_j\,$
being Boltzmann-Lorentz operator which acts onto $\,j$-th velocity.

Eventually we are interested in spatial distributions
\[
\,W_n(t,\bm{R})\,=\int
F_n(t,\bm{R},\bm{v}_1...\bm{v}_n)\,d\bm{v}_1...d\bm{v}_n\,
\]
In view of their definition and normalization of VDF, their
statistical sense is quite clear: $\,W_n(t,\bm{R})/N\,$ is density of
probability that at time moment $\,t>0\,$ an $\,n$-particle cluster
centered at point $\,\bm{R}\,$ contains (or at $\,n=1\,$ coincides
with) those unknown particle which at $\,t=0\,$ was localized near
point $\,\bm{R}=0\,$. Naturally, the larger is cluster the greater is
this probability. Equivalently, $\,W_1(t,\bm{R})\,$ is conditional
density of probability that well-known particle, at $\,t=0\,$
positioned near $\,\bm{R}=0\,$, later will be registered at given
point $\,\bm{R}\,$. In other words, $\,W_1(t,\bm{R})\,$ presents
probability distribution of Brownian displacement of any gas particle
during time interval $\,t\,$.

Of course, clusters (encounters) of $\,n>1\,$ particles are ephemeral
objects whose lifetime $\,\sim d/v_0\,$ is much smaller than mean
free flight time $\,\tau_f\,$. But they form constantly living ``gas
of encounters'' under description by second and higher of equations
(\ref{lod}). The latter have no principal difference from the first
equation since it also describes statistics of gas but not individual
particles.

\section{Roughened description of Brownian motion}
{\bf 1.\,} In the Boltzmannian kinetics, instead of (\ref{lod}) one
has the single equation for one-particle DF
$\,F_1(t,\bm{r},\bm{v})\,$,
\begin{equation}
\frac {\partial F_1}{\partial t}\,=\,-\,\bm{v}\cdot\frac {\partial
F_1}{\partial\bm{r}}+\nu\Lambda\, F_1\,\,\,,\label{bol}
\end{equation}
where the Boltzmann-Lorents operator $\,\Lambda\,$ tends velocity
distribution of gas particles to equilibrium one, $\,F_0(\bm{v})$. As
above, we can treat $\,F_1(t,\bm{r},\bm{v})\,$ as (small)
non-stationary non-uniform addition to uniform background due to a
priori information about initial position of one of particles. After
solving (\ref{bol}) with corresponding initial condition
$\,F_1(t=0,\bm{r},\bm{v})=W(\bm{r})F_0(\bm{v})\,$, the distribution
$\,W_1(t,\bm{r})=\int F_1(t,\bm{r},\bm{v})\,d\bm{v}\,$ will present
us information about a law of Brownian motion of the selected
particle.

Undoubtedly, in respect to large time scales, $\,t\gg \tau_f\,$, that
will be nothing but the Gaussian law, which can be expressed in
several equivalent forms:
\begin{eqnarray}
\lim_{s\rightarrow\infty} W_1(s^2t,s\bm{r})\,=\,(4\pi
Dt)^{-\mathcal{D}/2}\exp{(-\bm{r}^2/4Dt)} \,\,\,,\label{gl}\\
\lim_{s\rightarrow 0}
\,\Xi_1(t/s\,,\,\sqrt{s}\,\bm{k})\,=\,\exp{(-Dt\bm{k}^2)}
\,\,\,,\,\,\,\,\,\,\,\,\,\,\,\,\label{gcf}\\\, \lim_{s\rightarrow 0}
\,\,s\,\Phi_1(sp\,,\,\sqrt{s}\,\bm{k})\,=\,
1/(p+D\bm{k}^2)\,\,\,\,\,\,\,\,\,\,\,\,\,\,\,\,\label{glp}
\end{eqnarray}
Here $\,\Xi_1(t,\,\bm{k})\,$ is characteristic function of Brownian
motion and $\,\widetilde{\Xi}_1(p,\,\bm{k})\,$ its Laplace transform:
\begin{eqnarray}
\Xi_1(t,\,\bm{k})\,=\,\int\exp(i\,\bm{k}\cdot
\bm{r})\,W_1(t,\bm{r})\,d\bm{r}\,\,\,,\label{cf}\\
\Phi_1(p,\,\bm{k})\,=\,\int_0^\infty
\exp(-p\,t)\,\,\Xi_1(t,\,\bm{k})\,dt\,\,\label{lt}
\end{eqnarray}

Such long-time asymptotic is insensitive to concrete form of operator
$\,\Lambda\,$. Therefore, let us learn to obtain it in most simple
way, and then apply this our experience to equations (\ref{lod}).

The Boltzmann-Lorentz equation (\ref{bol}) directly yields
\begin{eqnarray}
\frac {\partial W_1(t,\bm{r})}{\partial t}\,=\,-\frac {\partial
\bm{J}_1(t,\bm{r})}{\partial\bm{r}}\,\,\,,
\,\,\,\,\,\,\,\,\,\,\,\,\,\,\,\,\,\,\,\,\,\,\label{dw}\\
\bm{J}_1(t,\bm{r})\,= \,\int
\bm{v}\,F_1(t,\bm{r},\bm{v})\,d\bm{v}\,=\,
\overline{\bm{v}}(t,\bm{r})\,W_1(t,\bm{r})\,\,\,,
\,\label{a}\\
\frac {\partial (\bm{J}_1(t,\bm{r}))_{\alpha}}{\partial t}=-\frac
{\partial }{\partial\bm{r}}\int \bm{v}v_{\alpha}\,F_1\,d\bm{v}+
\nu\int v_{\alpha}\Lambda F_1\,d\bm{v} \,\,\label{da}
\end{eqnarray}
Here, evidently, $\,\bm{J}_1(t,\bm{r})\,$ represents probability flow
and $\,\overline{\bm{v}}(t,\bm{r})\,$ conditional mean value of
velocity under given position of the particle.

Notice that soon after start there is some correlation between
current velocity of the particle and its displacement from initial
position. But then, at $\,t\gg\tau_f\,$, this correlation decreases
down to zero for any thermodynamically equilibrium, and hence
symmetrical, random walk (otherwise it would behave like ballistic
flight). In other words, the conditional mean velocity is much
smaller than above defined mean-square thermal velocity $\,v_0\,$:
\begin{eqnarray}
\overline{\bm{v}}(t,\bm{r})/v_0\,\rightarrow 0\,\label{vas}
\end{eqnarray}
for any possible realization of the walk (more precisely, one could
write
$\,\,\overline{\bm{v}}(t,\bm{r})W_1(t,\bm{r})/v_0W_1(t,0)\rightarrow
0\,\,$). Correspondingly, $\,F_1(t,\bm{r},\bm{v})\,$ factorizes into
$\,W_1(t,\bm{r})F_0(\bm{v})\,$ at $\,t\gg\tau_f\,$, and integral in
the first term of (\ref{da}) is dominated just by this factorized
asymptotic:
\begin{eqnarray}
\frac {\partial }{\partial\bm{r}}\int
\bm{v}v_{\alpha}\,F_1(t,\bm{r},\bm{v})\,d\bm{v}\,\rightarrow
\,v_0^2\,\frac {\partial }{\partial
r_{\alpha}}\,W_1(t,\bm{r})\,\,\label{vv}
\end{eqnarray}

In opposite, second integral in (\ref{da}), as well as the
probability flow $\,\bm{J}_1(t,\bm{r})\,$, is determined just by the
weak velocity-displacement correlation. Transform it as
\begin{eqnarray*}
\nu\int v_{\alpha}\Lambda F_1(t,\bm{r},\bm{v})\,d\bm{v}\,=\,\nu\int
F_1(t,\bm{r},\bm{v})\,[\Lambda^\dag\,
v_{\alpha}]\,d\bm{v}\,\,\,,
\end{eqnarray*}
where symbol $\,^\dag\,$ stands for transposition. In view of
time-reversal symmetry of collisions, $\,\Lambda^\dag\, v_{\alpha}\,$
is an odd function of $\,\bm{v}\,$. Therefore the latter integral is
an odd function of the conditional mean velocity
$\,\overline{\bm{v}}(t,\bm{r})\,$. At $\,t\gg\tau_f\,$, in view of
(\ref{vas}), is turns into linear function of
$\,\overline{\bm{v}}(t,\bm{r})\,$ and thus of
$\,\bm{J}_1(t,\bm{r})\,$. The coefficient follows from the asymptotic
factorization of $\,F_1\,$, and the result is
\begin{eqnarray}
\nu\int v_{\alpha}\Lambda F_1(t,\bm{r},\bm{v})\,d\bm{v}\,=
\,-\,\gamma\,(\bm{J}_1(t,\bm{r}))_{\alpha}\,\,\,,\label{g}\\
\gamma\,=\,-\frac {\nu}{\mathcal{D}} \int F_0(\bm{v})\left( \frac
{\partial }{\partial \bm{v}}\,\cdot \Lambda^\dag \bm{v}\right
)d\bm{v}\,\,\,\,\,\,\,\,\,\nonumber
\end{eqnarray}

Thus, velocity of particle is excluded from our consideration.
Combining (\ref{dw}), (\ref{da}), (\ref{vv}) with (\ref{g}), and
using $\,\nabla\,$ for spatial derivative, we come to approximate
equations which deal with particle displacement only:
\begin{eqnarray}
\frac {\partial W_1}{\partial t}\,=\,-\nabla\cdot
\bm{J}_1\,\,\,,\,\,\,\,\,\frac {\partial \bm{J}_1}{\partial
t}\,=\,-v_0^2\, \nabla\,W_1-\gamma\,\bm{J}_1\,\label{od}
\end{eqnarray}
Corresponding Laplace transform (\ref{lt}) of the characteristic
function (\ref{cf}) looks as
\begin{eqnarray}
\Phi_1(p,\,\bm{k})\,=\,\Xi_1(0,\,\bm{k})\left[p+\frac
{v_0^2\,\bm{k}^2}{p+\gamma }\right]^{-1}\label{glp1}
\end{eqnarray}
and evidently possesses Gaussian asymptotic (\ref{glp}), with
diffusivity $\,D=v_0^2/\gamma \,$. At that, velocity relaxation rate
$\,\gamma\,$ can be identified with $\,\tau_f^{-1}\,$.

Let us compare (\ref{glp1}) with formally exact answer when
$\,\Lambda\,$ is Fokker-Planck operator \cite{i1,i2} governing random
process with Gaussian velocity fluctuations (the Ornstein-Uhlenbeck
process),
\begin{eqnarray*}
\frac {\Phi_1(p,\,\bm{k})}{\Xi_1(0,\,\bm{k})}
\,=\,\left[p+v_0^2\,\bm{k}^2 \left[p+\gamma +\frac
{2v_0^2\,\bm{k}^2}{p\,+\,2\gamma
+...\,}\,\right]^{-1}\right]^{-1}\,\,
\end{eqnarray*}
This is infinite continued fraction with $\,nv_0^2\,\bm{k}^2\,$ in
nominators and $\,n\gamma\,$ in denominators ($\,n=1,2,...\,$). As
for equations (\ref{od}) and (\ref{glp1}), it can be shown that
formally they correspond to a random process with dichotomic velocity
fluctuations. However, from the point of view of large time scales
both random walks are identical. Realistic $\,\Lambda$'s account for
multi-state velocity hopping but lead to same random walks. This
example gives hope of that similar approach to system (\ref{lod})
will imitate its long-time behavior without principal losses.

{\bf 2.\,} In addition to distributions $\,W_n(t,\bm{R})\,$,
introduced in Sec.2, consider the ``flows''
\begin{eqnarray*}
\bm{J}_n(t,\bm{R})\,= \,\int
\bm{V}^{(n)}\,F_n(t,\bm{R},\bm{v}_1...\,\bm{v}_n)\,
d\bm{v}_1...d\bm{v}_n\,\,\label{an}
\end{eqnarray*}
In essence, equations (\ref{lod}) describe the same physics as
equation (\ref{bol}). Hence, all above reasonings about decay (at
$\,t\gg\tau_f\,$) of correlations between velocities and positions of
particles remain valid. And all they can be confirmed by direct
analysis of (\ref{lod}). Therefore, by analogy with (\ref{vv}) we can
write
\begin{eqnarray}
\nabla\int
\bm{V}^{(n)}V^{(n)}_{\alpha}\,F_n(t,\bm{R},\bm{v}_1...\,\bm{v}_n)
\,d^n\bm{v}\,\rightarrow \nonumber \\
\rightarrow \int
V^{(n)}_{\alpha}\bm{V}^{(n)}F_n^{\,eq}(\bm{v})\,d^n\bm{v}
\,\,\nabla\,W_n(t,\bm{R})\,=\nonumber \\=\, \frac
{v_0^2}{n}\,\,\nabla_{\alpha}\,W_n(t,\bm{r})\,\label{vvn}
\end{eqnarray}
(here and below $\,d^n\bm{v}\equiv d\bm{v}_1...\,d\bm{v}_n\,$ and
$\,\nabla\,$ replaces $\,\partial/\partial \bm{R}\,$). Analogue of
formula (\ref{g}) is
\begin{eqnarray}
\nu\int \bm{V}^{(n)}\sum_{j=1}^n\Lambda_j \,F_{n+1}
\,d^{n+1}\bm{v}\,= \nonumber \\
=\nu\int F_{n+1} \left[\frac 1n \sum_{j=1}^n\Lambda_j^\dag
\,\bm{v}_j\right]\,d^{n+1}\bm{v}=\nonumber\\
=\nu\int F_{n+1} \left[\frac {1}{n+1} \sum_{j=1}^{n+1}\Lambda_j^\dag
\,\bm{v}_j\right]\,d^{n+1}\bm{v}=\nonumber\\
=\,-\,\gamma\,\bm{J}_{n+1}(t,\bm{R})\,\label{gn}
\end{eqnarray}
with the same relaxation rate $\,\gamma\,$ as above. We took into
account that $\,\Lambda_j\,1\,=0\,$, therefore
$\,\Lambda_j\,\bm{v}_m\,=0\,$ at $\,j\neq m\,$, and that, naturally,
all $\,F_n(t,\bm{R},\bm{v}_1...\,\bm{v}_n)\,$ can be treated as fully
symmetric functions of velocities. Besides, now each of $\,F_n\,$
involves its own conditional mean velocity
$\,\overline{\bm{v}}_n(t,\bm{R})\,$. This derivation becomes quite
transparent in the case when $\,\bm{v}\,$ is eigenfunction of
$\,\Lambda\,$:\, $\,\Lambda\bm{v}=-\,\gamma \bm{v}\,$.

Consequently, we came to equations
\begin{eqnarray}
\frac {\partial W_n}{\partial t}\,=\,-\nabla\cdot
\bm{J}_n\,\,\,,\,\,\,\,\,\,\,\,\,\,\,\,\,\label{od1}\\
\frac {\partial \bm{J}_n}{\partial
t}\,=\,-\frac {v_0^2}{n}\, \nabla\,W_n
-\gamma\,\bm{J}_{n+1}\,\label{od2}
\end{eqnarray}
with initial conditions following from (\ref{is}):
\begin{equation}
W_n(t=0,\bm{R})\,=\,n\,W(\bm{R})\,\,\,, \,\,\,\bm{J}_n(t=0,\bm{R})=0
\label{isn}
\end{equation}
At that, since a shape of $\,W(\bm{R})\,$ will be of no importance,
we can simply put on $\,W(\bm{R})=\delta(\bm{R})\,$.

\section{Statistics of self-diffusion and low-frequency
fluctuations in diffusivity}

{\bf 1.\,} Applying to (\ref{od1})-(\ref{od2}) Fourier and Laplace
transforms, one can easy obtain formal solution to the problem
(\ref{od1})-(\ref{isn}) in the form of infinite series:
\begin{eqnarray}
p\,\Phi_n\,=\,n+\frac {D\bm{k}^2}{p}\sum_{s=n}^{\infty}
\prod_{m=n}^s\left[-\frac {\gamma}{p}\cdot \frac {m}{m+\gamma
D\bm{k}^2/p^2}\right]\,\label{sn}
\end{eqnarray}
Here $\,\Phi_n=\Phi_n(p,\bm{k})\,$ are defined analogously to
(\ref{lt}), and $\,D=v_0^2/\gamma\,$. In particular, at $\,n=1\,$ we
have
\begin{eqnarray}
p\,\Phi_1\,=\,1+\frac
{D\bm{k}^2}{p}\sum_{s=1}^{\infty}\,s\,B\left(s,1+\frac {\gamma
D\bm{k}^2}{p^2}\right) \left[-\frac {\gamma}{p}\right]^s\,\label{s1}
\end{eqnarray}
with $\,B(x,y)\,$ being beta-function. After that, using standard
integral representation of this function, we can transform the latter
series to compact integral form:
\begin{eqnarray}
p\,\Phi_1\,=\,1-Z\int_0^1\frac {(1-x)^Z}{(1+\gamma
x/p\,)^2}\,\,dx\,\,\,,\,\,\,\,\,Z\equiv \frac {\gamma
D\bm{k}^2}{p^2}\,\label{i1}
\end{eqnarray}
Let us discuss statistical properties of Brownian motion
corresponding to this formal result.

{\bf 2.\,} First, consider what is long-scale asymptotic of our
random walk, if interpret its asymptotic similarly to
(\ref{gl})-(\ref{glp}). In the limit like in (\ref{glp}), equation
(\ref{i1}) yields
\begin{eqnarray}
p\,\lim_{s\rightarrow 0}\,s\,\Phi_1(sp\,,\,\sqrt{s}\,\bm{k})\,=
\,\,\,\,\,\,\,\,\,\,\,\,\,\,\,\,\,\,\,\,\,\,\,\label{lsa}\\
=\,1-\frac {D\bm{k}^2}{p}\int_0^\infty \exp{\left(-\frac
{D\bm{k}^2}{p}\,x\right)}\,\frac {dx}{(1+x)^2}\,\nonumber\\
=\,1-\frac {D\bm{k}^2}{p}+\frac {(D\bm{k}^2)^2}{p}\int_0^\infty \frac
{e^{-\,x}}{D\bm{k}^2+p\,x}\,dx\nonumber
\end{eqnarray}
The last row is obtained with the help of integration by parts.
Integration by parts ``in opposite direction'' gives
\begin{eqnarray}
\lim_{s\rightarrow 0}\,s\,\Phi_1(sp\,,\,\sqrt{s}\,\bm{k})\,=
\,2\int_0^\infty \frac
{\exp{(-D\bm{k}^2\tau)}}{(1+p\,\tau)^3}\,d\tau\,\label{lsa1}
\end{eqnarray}
Performing here inverse Laplace transform, we obtain
\begin{eqnarray}
\lim_{s\rightarrow 0} \,\Xi_1(t/s\,,\,\sqrt{s}\,\bm{k})\,=
\,\,\,\,\,\,\,\,\,\,\,\,\,\label{xi}\\
=\,\int_0^\infty \exp{(-xD\bm{k}^2t\,)}\,\exp{\left(-\frac 1x
\right)} \,\frac {dx}{x^3}\,\nonumber
\end{eqnarray}
with $\,\Xi_1\,$ defined by (\ref{cf}). Then the inverse Fourier
transform of (\ref{xi}) produces
\begin{eqnarray}
\lim_{s\rightarrow\infty}\, W_1(s^2t,s\bm{R})\,=
\,\,\,\,\,\,\,\,\,\,\,\,\,\,\,\,\,\,\,\,\label{ngl}\\
\,\,\,=\,\frac {\Gamma (2+\mathcal{D}/2)}{(4\pi Dt)^{\mathcal{D}/2}\,
(1+\bm{R}^2/4Dt)^{2+\mathcal{D}/2}}\,\nonumber
\end{eqnarray}

This probability distribution describe random walk which has the same
characteristic scaling property $\,\bm{R}^2\sim t\,$ as Gaussian
random walk (Wiener process), but at that obeys extremely
non-Gaussian statistics. Formula (\ref{xi}) shows that this walk can
be represented as Wiener process with random but time-independent
diffusivity $\,\widetilde{D}\,$ distributed with probability density
$\,w(\widetilde{D}/D)/D\,$, where $\,w(x)=x^{-3}\exp{(-1/x)}\,$. That
is absolutely non-ergodic process. Besides, as it is evident from
(\ref{ngl}), the fourth and higher statistical moments of this
process are infinite. Apparently, non-Gaussian and non-ergodic
peculiarities of our random walk was too exaggerated by the
considered limiting procedure, and we have to return to equation
(\ref{i1}) which demonstrates that in fact all the moments are
finite.

{\bf 3.\,} Let $\,R(t)\,$ denotes projection of
$\,\mathcal{D}$-dimensional Brownian displacement $\,\bm{R(t)}\,$
during time $\,t\,$ onto some fixed direction, and vector
$\,\bm{k}\,$ is oriented in parallel to it. Since the direction is
fixed, we can treat $\,\bm{k}\,$ as scalar. Then compare expansions
of both sides of (\ref{i1}) into series over $\,k^2\,$:
\begin{eqnarray}
\Phi_1(p,k)-\frac 1p\,= \sum_{n=1}^\infty \frac
{(-k^2)^n}{(2n)!}\int_0^\infty e^{-p\,t}\,\langle R^{2n}(t)\rangle
\,dt\,=\nonumber \\
=\,\frac 1p\, \sum_{n=1}^\infty \left[-\frac {\gamma
Dk^2}{p^{\,2}}\right]^n\frac {1}{(n-1)!}\,\int_0^1\frac
{[-\ln{(1-x)}]^{\,n-1}}{(1+\gamma x/p\,)^2}\,dx\nonumber 
\end{eqnarray}
At $\,p/\gamma \ll 1\,$, which correspond to large time, the
comparison yields
\begin{eqnarray}
\int_0^\infty e^{-p\,t}\,\langle R^2(t)\rangle\,dt\,=\,\frac
{2D}{p^{\,2}}\,\,\,,\,\,\,\,\label{m2}\\
\int_0^\infty e^{-p\,t}\,\langle R^4(t)\rangle\,dt\, =\frac
{24D^2}{p^{\,\,3}}\,\ln{\left(1+\frac {\gamma}{p}\right)}\,\,\,,
\label{m4}\\
\int_0^\infty e^{-p\,t}\,\langle R^{\,2n}(t)\rangle\,dt\, =
\,\,\,\,\,\,\,\,\,\,\,\,\,\,\,\,\,\,\,\,\,\,\,
\,\,\,\,\,\,\,\,\,\,\,\,\,\,\,\,\,\,\,\,\nonumber\\
\,\,\,\,\,=\,(2n)!\,\,\zeta (n-1)\,\frac
{D^2\,v_0^{2n-\,4}}{p^{\,\,2n-\,1}}\,\,\,\,\,\,(n>2)\,\label{mn}
\end{eqnarray}
Here $\,\zeta (...)\,$ is the Riemann zeta function, and we used the
integrals
\begin{eqnarray}
\int_0^1\,\frac
{[-\ln{(1-x)}]^n}{x^2}\,dx\,=\,n!\,\zeta(n)\,
\,\,\,\,\,(n>1)\,\,,\nonumber\\
\,\,\,\, \int_0^1\,\frac {\ln{(1-x)}}{(1+ax)^2}\,dx\,=\,-\,\frac
{\ln{(1+a)}}{a(1+a)}\nonumber
\end{eqnarray}
Correspondingly to (\ref{m2})-(\ref{mn}),
\begin{eqnarray}
\langle R^2(t)\rangle\,=\,2Dt\,\,\,,\,\,\,\,\,\,\,\,\,\,\,\,\,\,
\,\,\,\,\,\,\,\,\,\,\,\,\,\,\,\,\,\,\,\label{t2}\\
\langle R^4(t)\rangle\, =\,12\,D^2t^2\,\ln{\gamma t}\,\,\,,
\,\,\,\,\,\,\,\,\,\,\,\,\,\,\,\,\,\,\,\,\,\label{t4}\\
\langle R^{\,2n}(t)\rangle\,=\,2n(2n-1)\,\zeta
(n-1)\times \,\,\,\,\,\,\,\,\,\nonumber\\
\times\,(Dt)^2\,(v_0t)^{2n-\,4}\,\,\,\,\,\,\,\,\,(n>2)\,\label{tn}
\end{eqnarray}
at $\,\gamma t\gg 1\,$. Thus, except mean square of the displacement,
all its higher-order moments are anomalously large from the point of
view of Gaussian random walk (which has $\,\langle R^{2n}(t)\rangle
=(2n)!!\,(2Dt)^n\,$).

Comparing formulas (\ref{t2})-(\ref{tn}) and distribution (\ref{ngl})
at $\,\mathcal{D}=1\,$, which also describes the projection
$\,R(t)\,$, we notice that these asymptotic formulas can be agreed
with probability distribution (\ref{ngl}) if the latter is
supplemented by its cut off at $\,|R(t)|\sim v_0t\,$. Hence, we can
improve (\ref{ngl}) if replace it, in asymptotical sense, by
\begin{eqnarray}
W_1(t,R)\approx\,\frac {\Gamma (5/2)\,G(R/v_0t)}{(4\pi Dt)^{1/2}\,
(1+R^2/4Dt)^{5/2}}\,\,\label{asw}
\end{eqnarray}
where $\,G(0)=1\,$ and $\,G(x)\,$ is extremely fast decaying function
at $\,x\rightarrow \pm\infty\,$ (this statement certainly follows
from very slow growth of coefficients on right-hand side of
(\ref{tn})). Thus we come to conclusions as follow.

(i)\, In respect to probabilistic characteristics of out random walk,
formula (\ref{asw}) practically coincides with (\ref{ngl}), that is
(\ref{ngl}) is quite satisfactory. However, in respect to higher
statistical moments, $\,\langle R^{\,2n}(t)\rangle\,$ with $\,n>1\,$,
there is not only strong quantitative difference between (\ref{ngl})
and (\ref{asw}) but also at least two principal differences.

(ii)\, The first difference is presented by fourth $\,\langle
R^{\,4}(t)\rangle\,$ moment. Such behavior of $\,\langle
R^{\,4}(t)\rangle\,$ as prescribed by (\ref{t4}) means that our
Brownian motion indeed can be represented as Gaussian one with random
diffusivity. However, diffusivity is not a ``static'' random value
(as (\ref{ngl}) would imply) but (as (\ref{asw}) implies) a random
process with characteristic low-frequency power spectrum $\,\sim
(\ln{f})^{\beta}/f \,$ (where $\,f\ll \gamma\,$ is frequency and
$\,\beta\,$ can be both positive and negative). In other words, the
diffusivity fluctuates like 1/f-noise. This theme was already
discussed not a once in \cite{i1,i2,bk1,bk2,bk3}.

(iii)\, The second peculiarity of our Brownian motion is presented by
moments $\,\langle R^{\,2n}(t)\rangle\,$ with $\,n>2\,$ which turn
out to be dependent on not only (average) diffusivity $\,D\,$ but
also on the thermal velocity $\,v_0\,$, i.e. characteristics of
separate free flights. This formal result can be interpreted as
evidence of that probability distribution of free flight time has a
power-law tail. Comparing (\ref{t4}) and moments corresponding to
purely free flight, $\,\langle R^{\,2n}(t)\rangle
=\,(2n)!!\,(v_0t)^{2n}\,$ (on $\,\langle R^{\,2n}(t)\rangle
=\,(v_0t)^{2n}\,$ for dichotomic velocity), we can suppose that
probability of anomalously long flights comparable with $\,t\,$
behaves as $\,\sim (Dt)^2/(v_0t)^4\,=\,1/(\gamma t)^2\,$\, (i.e. is
inversely proportional to square of mean number of collisions during
the observation time). But, strictly speaking, this issue needs in
investigation of many-time statistics of our Brownian motion. In any
case, on the whole it appears to be rather rather non-ergodic
process.

\section{Resume}

In this paper, previously constructed equations of the collisional
approximation to kinetics of statistically non-uniform gas were
applied to analysis of self-diffusion (Brownian motion) of particles
of infinite-volume gas in its thermodynamical equilibrium, under the
Boltzmann-Grad limit. With this purpose, a roughened version of the
equations was derived which allowed for their almost complete
solution. The latter demonstrated that statistics of the Brownian
motion is essentially non-Gaussian, not obeying the central limit
theorem, and hence essentially non-ergodic. It looks as if
diffusivity of the Brownian path had no definite value but instead
underwent low-frequency fluctuations with spectral properties similar
to that of 1/f-noise and with probability distribution possessing
power-law long tail. These results are in abrupt contradiction to
what follows from standard kinetic approximations, but agree with
Krylov's doubts \cite{kr} about obligatory ergodicity of mixing
dynamics.

\appendix
\section{Basic equations}

Our start point are the Liouville equation for $\,N\,$ particles in
container $\,0\leq r_{\alpha}\leq l\,$ ($\alpha =1\div\mathcal{D}$):
\begin{equation}
\frac {\partial F_N}{\partial t}\,=\,L^{(N)}F_N\,\,\,,\label{leq}
\end{equation}
\[
L^{(N)}\,\equiv\,-\sum_{j=1}^N \bm{v}_j\cdot\frac {\partial
}{\partial \bm{r}_j}\,+\sum_{1\leq j<m\leq N} L_{jm}
\,\,\,,\label{le}
\]
\[
L_{jm}\,\equiv\, -\nabla U(\bm{r}_j-\bm{r}_m)\cdot\left(\frac
{\partial }{\partial \bm{p}_j}-\frac {\partial }{\partial
p_m}\right)\,\,\,,
\]
with $\,U(\bm{r})\,$ being interaction potential, $\,\bm{p}_j
=M\bm{v}_j\,$ momentums, and
$\,F_N=F_N(t,\bm{r}_1...\bm{r}_N,\bm{v}_1...\bm{v}_N)\,$ normalized
probability density of phase point of the system, and the
corresponding Bogolyubov-Born-Green-Kirkwood-Yvon (BBGKY) equations
\begin{equation}
\frac {\partial F_1}{\partial t}=-\bm{v}_1\cdot\frac {\partial
F_1}{\partial r_1} +(N-1)\int_2 L_{12} F_2\,\,\,,\label{f1}
\end{equation}
\begin{equation}
\frac {\partial F_2}{\partial t}\,=\,L^{(2)}F_2+(N-2)\int_3
(L_{13}+L_{23}) F_3\,\,\,,\label{f2}
\end{equation}
\begin{equation}
\frac {\partial F_n}{\partial t}\,=\,L^{(n)}F_2+(N-n)\int_{n+1}
\sum_{j=1}^nL_{j\,n+1}F_{n+1}\,\,\,,\label{fn}
\end{equation}
where $\,\int_s...=\int...\,\,d\bm{v}_sd\bm{r}_s\,$ and
$\,F_n=\int_{n+1}F_{n+1}\,$. In the limit of infinitely big
container, $\,N\rightarrow\infty$, $\, l\rightarrow\infty\,$, $\,\nu
=N/\Omega=$\,const\, ($\,\Omega \equiv l^{\mathcal{D}}$), one must
consider non-normalized distribution functions $\,\Omega^nF_n\,$. If
designate them again as $\,F_n\,$ then (\ref{fn}) turns into
\begin{equation}
\frac {\partial F_n}{\partial t}\,=\,L^{(n)}F_2+\nu\int_{n+1}
\sum_{j=1}^nL_{j\,n+1}F_{n+1}\,\,\label{fnn}
\end{equation}

The equations (\ref{leq})-(\ref{fn}) hold also for particles on torus
with equal dimensions, $\,0\leq r_{\alpha}\leq l\,$, if only
supplement them with the periodic boundary conditions and replace
$\,U(\bm{r})\,$ by $\,U(\bm{\rho})\,$, where $\,\rho =\rho (r)\,$ is
factual distance between two particles on torus:
$\,\rho_{\alpha}=r_{\alpha}\,$ when $\,|r_{\alpha}|\leq l/2\,$ and
$\,\rho_{\alpha}=-$sign$(r_{\alpha})(l-|r_{\alpha}|)\,$ when
$\,|r_{\alpha}|> l/2\,$. However, if we want to consider unbounded
Brownian motion of particles along the torus, we should introduce
more rich distribution functions which keep count of their rotations.
This is subject of second part of the paper.

\section{Collisional approximation}

Assuming that our gas is sufficiently rarefied, that is $\,\delta
/a\ll 1\,$, with $\,a=l/N^{1/\mathcal{D}}\,$ being typical distance
between neighbour particles and $\,\delta\,$ interaction radius, we
want to describe it approximately in terms of Boltznmannian pair
collision operators (``collision integrals''). With this purpose,
basing on the exact BBGKY equations (see Appendix A) consider
$\,F_2(t,\bm{r}_1,\bm{r}_2,\bm{v}_1,\bm{v}_2)\,$ at
$\,|\bm{r}_2-\bm{r}_1|<d\,$ where $\,\delta\ll d\ll a\,$. Any such
configuration can be interpreted as a snapshot of certain process of
collision of two particles or, may be, their close encounter without
significant interaction. Therefore let us name the region
$\,|\bm{r}_2-\bm{r}_1|<d_2\,$ ``{\it collision box}'' and
characterize states inside it by inner time $\,\Theta\,$ of
corresponding collision plus relative velocity and
$\,(\mathcal{D}-1)$-dimensional impact parameter at its incoming
stage. We can choose $\,\Theta =0\,$ at perigee of the collision or
encounter. If using these new variables, (\ref{f2}) can be rewritten
in the form
\begin{equation}
\frac {\partial F_2}{\partial t}\,=-\bm{V}^{(2)}\cdot \frac {\partial
F_2}{\partial \bm{R}}-\frac {\partial F_2}{\partial
\Theta}+(N-2)\int_3 (L_{13}+L_{23}) F_3 \label{f22}
\end{equation}
with $\,\bm{R}=(\bm{r}_1+\bm{r}_2)/2\,$ and
$\,\bm{V}^{(2)}=(\bm{v}_1+\bm{v}_2)/2\,$ being position and velocity
of center of gravity of the pair. At that, by definition of the inner
time,
\begin{equation}
-\frac {\partial F_2}{\partial \Theta}\,=\,-\,\bm{v}_{12}\cdot \frac
{\partial F_2}{\partial \bm{r}_{12}} + L_{12}F_2\,\,\,, \label{teta}
\end{equation}
where $\,\bm{r}_{12}=\bm{r}_2-\bm{r}_1\,$ and
$\,\bm{v}_{12}=\bm{v}_2-\bm{v}_1\,$.

Any dependence of $\,F_2\,$ on $\,\Theta\,$, at least at $\,|\Theta
|\,\sim \,$ typical duration of collisions $\,\tau_c\,$, does mean
that at statistical ensemble under consideration a number (or
probability of realization) of pairs coming into ($\,\Theta <0\,$) a
given sort of collisions differs from a number (or probability of
realization) of pairs leaving ($\,\Theta >0\,$) the same collision.
Generally such is the case, and such difference obstructs collisional
approximation, because collision operator must conserve probability
(just as collision conserves number of particles). Or alternatively
we should neglect this difference and postulate that
\begin{equation}
\frac {\partial F_2}{\partial
\Theta}\,=\,0\,\,\,\,\,\,\text{when}\,\,\,\,\bm{r}_{12}\,\in\,
collision\,\,box\,\, \label{hyp}
\end{equation}
in (\ref{f22}), thus ensuring conservation of probability in the
course of collisions. The destination of (\ref{hyp}) is correct
transition from separate dynamic states to coherent collision as the
whole.

According to (\ref{teta}), relative displacement of colliding
particles is included into collision. Consequently, it drops out of
outward characterization of collision as the whole, and the latter
obeys the equation
\begin{equation}
\frac {\partial F_2}{\partial t}\,=-\bm{V}^{(2)}\cdot \frac {\partial
F_2}{\partial \bm{R}}+(N-2)\int_3 (L_{13}+L_{23}) F_3 \label{f222}
\end{equation}
At that, equation (\ref{teta}), which is probabilistic equivalent of
Hamilton dynamics inside {\it collision box},
\begin{equation}
\frac {d \bm{r}_{12}}{d
\Theta}\,=\,\bm{v}_{12}\,\,\,\,,\,\,\,\,\,M\,\frac {d \bm{v}_{12}}{d
\Theta}\,=\,-\frac {\partial U(\bm{r}_{12})}{\partial
\bm{r}_{12}}\,\,\,, \label{ham}
\end{equation}
together with (\ref{hyp}) serves for transformation of last term in
(\ref{f1}) into usual Boltzmann collision integral \cite{i1,i2}.

Here we detect principal difference from Boltzmann's theory: equation
(\ref{f222}) certainly forbids Boltzmann's Stosshalansatz (except
trivial situation when $\,F_1\,$ is exactly spatially uniform). {\it
``Molecular chaos'' fails in collision box, just where it was most
wanted. However, nothing prevents ``molecular chaos'' outside
collision box, and it really holds there (at least in the
Boltzmann-Grad limit).}

Clearly, the function $\,F_2\,$ governed by equation (\ref{f222})
presents probability measure (ensemble-average density) of one or
another sort of collisions (differentiated e.g. by their impact
parameters and relative velocity $\,\bm{v}_{12}\,$). Notice that it
becomes normalized to unit, in respect to $\,\bm{R}^{(2)}\,$ and both
velocities, if multiply $\,F_2\,$ by gas volume $\,\Omega\,$. The
first term on right-hand side of (\ref{f222}) describes drift of
collisions in spatially non-uniform gas while the second their
redistribution over sorts due to intervention of a third outer
particle.

Hence, next we should consider clusters of three close particles
belonging to a reasonable ``collision box'', e.g.
$\,|\bm{r}_i-\bm{r}_j|<d\,$, and again connect instant dynamic states
inside it into 3-particle collision as a single whole. Of course,
that is not literally 3-particle collision but sooner 3-particle
encounter with possibility of pair collisions among some of the
particles. Again, distribution of such 3-particle events over their
sorts can change under influence by a fourth exterior particle.
Continuing such scheme to $\,n$-particle encounters, we decompose the
$\,n$-particle Liouville operator similarly to (\ref{f22}):
\begin{equation}
L^{(n)}F_n\,=-\bm{V}^{(n)}\cdot \frac {\partial F_n}{\partial
\bm{R}}-\frac {\partial F_n}{\partial \Theta}\,\,\,, \label{ln}
\end{equation}
with $\,\bm{R}=(\bm{r}_1+...+\bm{r}_n)/n\,$,
$\,\bm{V}^{(n)}=(\bm{v}_1+...+\bm{v}_n)/n\,$, and $\,\Theta\,$ being
inner time of an encounter. Then again it is necessary to postulate
conservation of probability in collision box:
\begin{equation}
\frac {\partial F_n}{\partial
\Theta}\,=\,0\,\,\,\,\,\,\text{when}\,\,\,\,\bm{r}_i-\bm{r}_j\,\in\,
collision\,\,box\,\,\,, \label{hypn}
\end{equation}
which simplifies (\ref{fn}) into equation
\begin{equation}
\frac {\partial F_n}{\partial t}=-\bm{V}^{(n)}\cdot \frac {\partial
F_n}{\partial \bm{R}}\,+(N-n)\int_{n+1}\sum_{j=1}^n L_{j\,n+1}F_{n+1}
\label{fnn}
\end{equation}
similar to (\ref{f222}). At the same time, (\ref{hypn}) helps to
transform the last term in similar equation for $\,F_{n-1}\,$ into
sum of $\,n-1\,$ collision integrals.

Thus we come to a chain of coupled kinetic equations:
\begin{equation}
\frac {\partial F_n}{\partial t}\,=\,-\bm{V}^{(n)}\cdot\frac
{\partial F_n}{\partial \bm{R}}+(N-n)\sum_{j=1}^n
S_{j\,n+1}F_{n+1}^{\,in}\,\,\,\, \label{ch}
\end{equation}
where $\,n=1...\,N\,$\, and $\,S_{j\,n+1}\,$ is Boltzmannian
collision operator describing collision between $\,j$-th particle
from the $\,n$-particle cluster and one more ``$\,(n+1)$-th''
particle from its exterior. Functions $\,F_{n+1}^{\,in}\,$ represent
$\,F_{n+1}\,$ at boundary of $\,(n+1)$-particle collision box, more
precisely, at those part of the boundary which correspond to incoming
stage of these collisions. At that, differences $\,\sim d <a\,$
between centers of gravity of successive collision boxes are
neglected because of smallness in comparison with the mean free path
$\,\lambda\sim a\,(a/\delta)^{\mathcal{D}-1}\,$. In the limit of
infinite gas, after normalization to its volume, $\,\Omega^n
F_n\rightarrow F_n\,$, equations (\ref{ch}) turn to infinite chain of
equations from \cite{i1,i2},
\begin{equation}
\frac {\partial F_n}{\partial t}\,=\,-\bm{V}^{(n)}\cdot\frac
{\partial F_n}{\partial \bm{R}}+\nu\sum_{j=1}^n
S_{j\,n+1}F_{n+1}^{\,in}\,\,\,\, \label{inf}
\end{equation}

Importantly, the boundary functions $\,F_{n+1}^{\,in}\,$ in
(\ref{ch}) and (\ref{inf}) in no way are ruled by these equations
themselves. In essence, $\,F_{n+1}^{\,in}\,$ supply initial
conditions to equations (\ref{ham}), as well as similar
$\,n$-particle equations, and must be determined independently, by
lacing $\,F_{n+1}$'s with probability distributions at exterior of
the collision box. But its boundary is just border between violation
and observance of ``molecular chaos'' (see Italic text above).
Therefore, here somewhat mediate must be expected. For example,
\[
F_2^{\,in}(t,\bm{R},\bm{v}_1,\bm{v}_2)\,=\,
F_1^{\,\prime}(t,\bm{R},\bm{v}_1)F_1^{\,\prime\prime}(t,\bm{R},\bm{v}_2)
\,\,
\]
where generally neither $\,F_1^{\,\prime}(t,\bm{R},\bm{v})\,$ nor
$\,F_1^{\,\prime\prime}(t,\bm{R},\bm{v})\,$ coincide with the
one-particle distribution $\,F_1(t,\bm{R},\bm{v})\,$. Thus
Stosshalansatz is valid with respect to velocities of colliding
particles only but not to their positions ! More concrete formulation
of such ``weakened molecular chaos'' follows again from conditions of
conservation of probability (now at the border of ``molecular
chaos'') \cite{i1,i2}:
\begin{eqnarray}
F_{n+1}^{in}(t,\bm{R},\bm{v}_1...\bm{v}_n,
\bm{v}_{n+1})\,=\,\widetilde{F}_1(t,\bm{R},\bm{v}_{n+1})\,
\times\nonumber \\
\times\,\int F_{n+1}(t,\bm{R},\bm{v}_1...\bm{v}_n, \bm{v}^{\,\prime
})\,d\bm{v}^{\,\prime }\,\,\,\,,\,\,\,\,\,\,\,\,\,\,\,\,\label{wch}
\end{eqnarray}
where $\,\widetilde{F}_1\,$ is local one-particle velocity
distribution,
\begin{equation}
\widetilde{F}_1(t,\bm{R},\bm{v})\,\equiv\,
\frac{F_1(t,\bm{R},\bm{v})}{\int F_1(t,\bm{R},\bm{v}^{\,\prime
})\,d\bm{v}^{\,\prime }} \,\,\,,\label{w1}
\end{equation}
normalized to unit.

But what about $\,\mathcal{D}(n-1)-1\,$ those arguments of $\,F_n\,$
at $\,n>1\,$ (in particular, impact parameters) which concretize
geometry of $\,n$-particle encounters but not written out above ? If
we made attempt to take them into account, our theory would be not
simpler than underlying BBGKY equations. Therefore we suppose that
$\,F_n\,$ in (\ref{ch}) and (\ref{inf}) represent distributions
averaged over all sorts of the encounters. This roughening of the
theory looks be caricature, but it can be legalized in many respects
under the Boltzmann-Grad limit ($\,\nu\rightarrow\infty\,$,
$\,\delta\rightarrow 0\,$, $\,\lambda\sim 1/\nu\delta^{\mathcal{D}-1}
=\,$const\,). What is the chief thing, at that the theory keeps
principal open-chain structure of its equations. By this reasons, in
Sec.2 we venture to roughen the theory yet stronger.

Additional simplification appears when we want to investigate
Brownian motion (self-diffusion) of gas particles in absence of their
collective hydrodynamic motion, that is in thermodynamically
equilibrium gas. This purpose can be achieved if apply the trick well
known in the classical kinetic theory of gases \cite{rdl}, namely, if
treat ``outside $(n+1)$-th particle'' in (\ref{wch})-(\ref{w1}) as a
particle of equilibrium thermostat:
$\,\widetilde{F}_1(t,\bm{R},\bm{v})\rightarrow F_0(\bm{v})\,$, where
$\,F_0(\bm{v})\,$ is equilibrium velocity distribution (standard
Gaussian at $\,N\rightarrow\infty\,$). Correspondingly, the Boltzmann
collision operator $\,S_{j\,n+1}\,$ transforms into the
Boltzmann-Lorentz operator \cite{rdl}, $\,\Lambda_j\,$, defined by
\begin{eqnarray*}
S_{j\,n+1}\,[F(...\bm{v}_j...)W_0(\bm{v}_{n+1})]\,=\,\Lambda_j\,
F(...\bm{v}_j...)\\
\end{eqnarray*}
Equations (\ref{inf})-(\ref{wch}) take the form \cite{i1,i2}
\begin{equation}
\frac {\partial F_n}{\partial t}\,=\,-\bm{V}^{(n)}\cdot\frac
{\partial F_n}{\partial \bm{R}}+\nu\sum_{j=1}^n \Lambda_j\int
F_{n+1}\,d\bm{v}_{n+1}\, \label{lo}
\end{equation}
and similarly equations (\ref{ch}) do change. Thus we find ourselves
in situation when spatial non-uniformity of distribution functions
$\,F_n\,$ has purely statistical origin: it describes not a
thermodynamic perturbation but information about positions of some
particles only (even may be a single particle).



\begin{thebibliography}{28}

\bibitem{i1}
Yu.\,E.\,Kuzovlev, ``Bogolyubov-Born-Green-Kirkwood-Yvon equations,
self-diffusion and 1/f-noise in a slightly nonideal gas'',
Sov.Phys.-JETP, {\bf 67} (12), 2469 (1988) [in Russian:
Zh.Eksp.Teor.Fiz., {\bf 94}, No.12, 140 (1988)].

\bibitem{i2}
Yuriy\,E.\,Kuzovlev, ``Kinetical theory beyond conventional
approximations and 1/f-noise'', arXiv: cond-mat/9903350.

\bibitem{bk1}
Yu.\,E.\,Kuzovlev and G.\,N.\,Bochkov, ``On the origin and
statistical characteristics of 1/f-noise'', ``Radiophysics and
Quantum Electronics'', No.3 (1983) [in Russian:
Izv.\,VUZov.-Radiofizika, {\bf 26}, 310 (1983)].

\bibitem{bk2}
G.\,N.\,Bochkov and Yu.\,E.\,Kuzovlev, ``On the probabilistic
characteristics of 1/f-noise'', ``Radiophysics and Quantum
Electronics'', No.9 (1984) [in Russian: Izv.\,VUZov.-Radiofizika,
{\bf 27}, 1151 (1984)].

\bibitem{bk3}
G.\,N.\,Bochkov and Yu.\,E.\,Kuzovlev, ``New in 1/f-noise studies'',
Sov.Phys.-Usp., {\bf 26}, 829 (1983) [in Russian: UFN, {\bf 141}, 151
(1983)].

\bibitem{an}
D.\,V.\,Anosov et al., ``Dynamical systems with hyperbolic
behaviour'', in ''Modern problems of mathematics. Fundamentals. Vol.
66'', ed. R.V.Gamkrelidze, VINITI, Moscow, 1991 (in Russian).

\bibitem{cy}
N.\,Chernov and L.-S.\,Young, ``Decay of correlations for Lorentz
gases and hard balls'', in ``Hard Ball Systems and the Lorentz Gas'',
ed. D.\,Szasz, Encyclopaedia of Mathematical Sciences {\bf 101}, pp.
89-120, Springer, 2000 (http://www.math.uab.edu/chernov).

\bibitem{ns}
N.\,Sim\'{a}nyi, ``The Boltzmann-Sinai ergodic hypothesis in full
generality'', arXiv: math.DS/0510022.

\bibitem{rdl}
P.\,Resibois and M.\,de\,Leener. Classical kinetic theory of fluids.
Wiley, New-York, 1977.

\bibitem{kr}
N.\,S.\,Krylov. Works on the foundation of statistical physics.
Princeton Univ. Press, Princeton, 1979.

\end{thebibliography}


\end{document}